\documentclass[pra,aps,preprint,showpacs]{revtex4}

\usepackage{graphicx}
\usepackage{amssymb} 

\begin{document}
\title{Witnessing negativity of Wigner function by estimating fidelities of cat-like states from homodyne measurements}
\author{Jarom\'{\i}r Fiur\'{a}\v{s}ek and Miroslav Je\v{z}ek}
\affiliation{Department of Optics, Faculty of Science, Palack\'{y} University, 17. listopadu 12,
77146 Olomouc, Czech Republic}

\begin{abstract}
We derive sampling functions for estimation of quantum state fidelity with Schr\"{o}dinger cat-like states, which are defined
as superpositions of two coherent states with opposite amplitudes. We also provide sampling functions for fidelity with squeezed Fock states 
that can approximate the cat-like states and can be generated from Gaussian squeezed states by conditional photon subtraction. The fidelities can be determined
by averaging the sampling functions over quadrature statistics measured by homodyne detection. The sampling functions are designed such that they 
can compensate for losses and inefficient homodyning provided that the overall efficiency exceeds certain threshold. 
The fidelity with an odd coherent state and the fidelity with a squeezed odd Fock state provide convenient witnesses 
of negativity of Wigner function of the measured state. The negativity of Wigner function at the origin of phase space is certified if any of these 
fidelities exceeds $0.5$. Finally, we discuss the possibility of reducing the statistical uncertainty of the fidelity estimates 
by a suitable choice of the dependence of the number of quadrature samples on the relative phase shift between local oscillator and signal beam.

\end{abstract}


\maketitle

\section{Introduction}

Negativity of Wigner function represents one of the most striking signatures of nonclassicality of the quantum states of light. 
The quantum states exhibiting negative Wigner function are not only of fundamental interest but they also represent a valuable resource for quantum enhanced metrology \cite{Giovanetti11}, 
 optical quantum information processing \cite{Furusawa11,Pan12} and tests of Bell inequality violation with balanced homodyning \cite{Wenger03,Nha04,Patron04}. 
Very recently, it was shown that states with negative Wigner function are necessary for quantum enhanced computing because quantum circuits involving 
only states and operations with positive Wigner functions can be classically efficiently simulated \cite{Mari12,Veitch12,Veitch13}.

Given the fundamental importance and potential practical applications of highly non-classical states of light with negative Wigner function, 
their experimental generation  has attracted a great deal of attention during recent years. 
The states of propagating light fields exhibiting negative Wigner function can be prepared from initial Gaussian squeezed states 
by exploring quantum correlations between several optical modes and conditioning on the single photon detection on some of the modes. Depending on the chosen configuration one can either conditionally prepare a Fock state 
\cite{Lvovsky01,Ourjoumtsev06fock} or perform a conditional  single-photon addition \cite{Zavatta04,Barbieri10} or subtraction \cite{Ourjoumtsev06,Neergaard06,Takahashi08,Gerrits10,Neergaard10}. These latter operations proved to be extremely powerful and useful in engineering various quantum states and operations \cite{Usuga10,Neergaard10qubit,Marek10,Zavatta11,Tipsmark11,Blandino12}. In particular, by conditionally subtracting a single photon from a squeezed vacuum state one can generate a squeezed single-photon state that closely approximates a coherent superposition of two coherent states, also referred to as Schr\"{o}dinger cat-like state. 
 
The cat-like states represent a crucial resource for quantum computing with coherent states, where qubits are encoded into superpositions of two coherent states with sufficiently small overlap
\cite{Jeong02,Ralph03,Lund08}. The advantage of this scheme is that the two-qubit entangling gate, which is difficult to implement for single-photon qubits, can be implemented simply by interference on a beam splitter. On the other hand, the single-qubit Hadamard gate becomes challenging in this framework, because it amounts to generation of coherent superpositions from input coherent states. A scalable implementation of the Hadamard gate as well as the whole quantum computing scheme requires quantum teleportation where auxiliary cat-like states serve as quantum channels.

In practice, any state preparation procedure is necessarily imperfect and influenced by losses, noise, and other decoherence effects. It is thus very important to reliably 
characterize the generated states and certify their desired properties such as the negativity of Wigner function. In a vast majority of current experiments on generation of cat-like states of traveling light fields, this characterization is accomplished by homodyne detection on the prepared state. By varying the phase between local oscillator and signal beam, statistics of various rotated quadratures is measured, which provides a tomographically complete set of data \cite{Vogel89} from which the state is subsequently reconstructed. 

Here we consider scenario where we are interested only in certain specific characteristics of the state. In particular, we will investigate determination of the fidelity of the measured state with a cat-like state from the homodyne data. The fidelity provides a convenient and succinct characterization of the quality of the prepared state and we show that it can serve as a witness of the negativity of Wigner function of the prepared state. We will derive sampling functions \cite{Leonhardt97,Welsch99} that enable fast and direct estimation of the cat-state fidelity by averaging the sampling function over the measured quadrature statistics. This approach is appealing because it 
avoids the full tomographic reconstruction of the state from the measured data which may be computationally demanding. 
In addition, the resulting fidelity estimator is linear hence its statistical error can be easily determined.

The rest of the paper is organized as follows. In Section II we briefly review the mathematical model of homodyne detection and derive sampling function for fidelity with coherent superposition of two coherent states. We also show that the fidelity with an odd coherent state provides an upper bound on the value of the Wigner function at the origin of phase space, so it can be used to witness the negativity of $W$ at this point. In Section III we derive sampling functions for diagonal density matrix elements in the basis of squeezed Fock states, which represent fidelities of the measured state with the squeezed Fock states. In Section IV we investigate reduction of statistical uncertainty of the estimated fidelities for a fixed total number of measurements by an inhomogeneous sampling where the number of quadrature measurements depends on the phase between signal and local oscillator beams. Finally, Section V contains a brief summary and conclusions.

\section{Sampling functions for cat-state fidelity}

A homodyne detector measures the rotated quadrature $x_\theta=x\cos\theta+p\sin \theta$, where $x$ and $p$ denote the amplitude and phase quadratures
that satisfy canonical commutation relations $[x,p]=i$ \cite{Leonhardt97}. A realistic homodyne detector with detection efficiency $\eta <1$ samples probability distribution $w(x_\theta';\theta,\eta)$ 
of a noisy quadrature
\begin{equation}
x_\theta^\prime=\sqrt{\eta}x_\theta+\sqrt{1-\eta}x_{\mathrm{vac}},
\label{xthetaprime}
\end{equation}
where $x_{\mathrm{vac}}$ denotes quadrature operator of an auxiliary vacuum state. 
Our goal is to estimate from the homodyne data the fidelity of the measured quantum state $\rho$ with a cat-like state represented by a coherent superposition of two coherent states,
\begin{equation}
F_C= \langle \psi_C| \rho |\psi_C\rangle,
\label{FC}
\end{equation}
where
\begin{equation}
|\psi_C\rangle=\frac{1}{\sqrt{2\mathcal{N}}}\,\left(|\alpha\rangle+e^{i\phi}|-\alpha\rangle\right),
\label{catstate}
\end{equation}
$|\alpha\rangle$ denotes a coherent state with  amplitude $\alpha$ and $\mathcal{N}=1+e^{-2|\alpha|^2}\cos\phi$ is a normalization factor.
More specifically, we seek a sampling function $S_F(x_\theta',\theta;\eta,\alpha,\phi)$ whose average over the measured quadrature statistics yields the fidelity (\ref{FC}),
\begin{equation}
F_C=\frac{1}{2\pi}\int_{0}^{2\pi} \int_{-\infty}^\infty w(x_\theta';\theta,\eta)S_F(x_\theta',\theta;\eta,\alpha,\phi) \, d x_\theta' \, d\theta.
\end{equation}
If we insert the explicit formula (\ref{catstate}) for the cat-like state into Eq. (\ref{FC}), we get
\begin{eqnarray}
 F_C&=&\frac{1}{2\mathcal{N}}\left(\langle \alpha|\rho|\alpha\rangle+\langle -\alpha|\rho|-\alpha\rangle\right) 
 +\frac{1}{2\mathcal{N}}\left(e^{i\phi}\langle \alpha|\rho|-\alpha\rangle+e^{-i\phi}\langle -\alpha|\rho|\alpha\rangle\right).
\end{eqnarray}
This formula indicates that the fidelity $F_C$ is closely related to the Husimi Q-function of the state $\rho$, defined as \cite{Walls94}
\begin{equation}
Q(\alpha,\alpha^\ast)=\frac{1}{\pi}\langle \alpha |\rho|\alpha\rangle = \frac{1}{\pi} e^{-\alpha\alpha^\ast} \sum_{m,n=0}^\infty \frac{\alpha^{\ast m}\alpha^n}{\sqrt{m!\,n!}} \, \rho_{m,n},
\end{equation}
where $\rho_{m,n}=\langle m|\rho|n\rangle$ denotes density matrix element in Fock basis. 
If we formally assume that $\alpha$ and $\alpha^\ast$ are completely independent variables then we have for instance $\pi Q(-\alpha,\alpha^\ast)=e^{2\alpha\alpha^\ast} \langle\alpha|\rho|-\alpha\rangle$, and we can thus write
\begin{eqnarray}
 F_C=\frac{\pi}{2\mathcal{N}}\left[ Q(\alpha,\alpha^\ast)+Q(-\alpha,-\alpha^{\ast})  \right]+ 
\frac{\pi e^{-2|\alpha|^2}}{2\mathcal{N}}\left[ e^{i\phi}Q(-\alpha,\alpha^\ast)+e^{-i\phi}Q(\alpha,-\alpha^{\ast})  \right].
\label{FCQ}
\end{eqnarray}
The sampling function for fidelity $F_C$ can therefore be constructed from a sampling function $S_Q$ for the Husimi Q-function \cite{D'Ariano99,Richter99,Richter00,Fiurasek01}.
For the sake of completeness we briefly recapitulate the derivation of $S_Q$ since in the next section 
we will extend this procedure to derive a sampling function for the Q-function of a squeezed copy of the state. 
The Q-function can be calculated as an inverse Fourier transform of the characteristic function 
$C_{\mathcal{A}}(\beta,\beta^\ast)=e^{-|\beta|^2/2} \left\langle \exp(\beta a^\dagger-\beta^{\ast}a ) \right\rangle$ 
corresponding to antinormally ordered moments of annihilation and creation operators,
\begin{equation}
Q(\alpha,\alpha^\ast)=\frac{1}{\pi^2 } \int _{-\infty}^\infty \int _{-\infty}^\infty C_{\mathcal{A}}(\beta,\beta^\ast) \, e^{\alpha\beta^\ast-\alpha^\ast\beta}\,d^2\beta.
\label{QFourier0}
\end{equation}

By introducing polar coordinates for the complex variable $\beta=i k e^{i\theta}$ we obtain a relation between the characteristic function $C_{\mathcal{A}}$ and 
a characteristic function of rotated quadrature operator $x_{\theta}$,
$C_{\mathcal{A}}(\beta,\beta^\ast)=e^{-k^2/2} \left\langle \exp(i\sqrt{2}k x_{\theta}) \right\rangle$. If we insert this expression into Eq. (\ref{QFourier0}) we get
\begin{equation}
Q(\alpha,\alpha^\ast)=\frac{1}{\pi^2} \int_0^{2\pi} \int_0^\infty k \,e^{-k^2/2}\left\langle e^{\sqrt{2}i k x_\theta}\right \rangle e^{-\sqrt{2} i k \tilde{x}_\theta} 
\,\mathrm{d}k\, \mathrm{d}\theta,
\label{Qfunction}
\end{equation}
where $\tilde{x}_{\theta}=\frac{1}{\sqrt{2}}(\alpha e^{-i\theta}+\alpha^{\ast} e^{i\theta})$.
Using  Eq. (\ref{xthetaprime}) we can establish a relationship between the characteristic function of the measured quadrature $x_\theta'$ and the characteristic function of quadrature $x_\theta$
which appears in the formula (\ref{Qfunction}). Since the quadrature  $x_{\mathrm{vac}}$  is not correlated with $x_\theta$ 
and exhibits Gaussian probability distribution with zero mean and variance $\frac{1}{2}$, the characteristic function of $x_\theta'$ becomes a product of characteristic functions of
$x_\theta$ and $x_{\mathrm{vac}}$,
\begin{equation}
\langle e^{i\xi x_{\theta}^\prime}\rangle= \langle e^{i\sqrt{\eta}\xi x_{\theta}}\rangle \, e^{-(1-\eta)\xi^2/4}.
\label{charfunctionrelation}
\end{equation}
We set $\xi=k\sqrt{2/\eta}$, express the characteristic function of $x_\theta'$ as a Fourier transform of $w(x_\theta';\theta,\eta)$, and after some algebra we obtain
\begin{equation}
\langle e^{\sqrt{2}i k x_\theta} \rangle=e^{\frac{1-\eta}{2\eta}k^2}\int_{-\infty}^{\infty}w(x_\theta';\theta,\eta)\,e^{i \sqrt{2} k x_{\theta}'/\sqrt{\eta}} \,d x_\theta'.
\label{charxtheta}
\end{equation}
If we insert the expression (\ref{charxtheta}) into Eq. (\ref{Qfunction}) and evaluate the integral over $k$ then we arrive at
\begin{equation}
Q(\alpha,\alpha^\ast)=\frac{1}{2\pi}\int_{0}^{2 \pi} \int_{-\infty}^\infty w(x_\theta';\theta,\eta)S_Q(x_\theta',\theta;\eta,\alpha,\alpha^\ast) \, \mathrm{d}x_\theta' \, \mathrm{d}\theta,
\end{equation}
where the sampling function for the Husimi Q-function reads \cite{D'Ariano99,Richter99,Richter00}
\begin{equation}
S_Q(x_\theta',\theta;\eta,\alpha,\alpha^\ast)=\frac{2}{\pi}\frac{\eta}{2\eta-1}\left[1-\sqrt{\pi}y e^{-y^2}\mathrm{erfi}(y)\right].
\label{SQformula}
\end{equation}
Here $\mathrm{erfi}(z)=-i\, \mathrm{erf}(i z)$ denotes the error function of imaginary argument, and  
\begin{equation}
y=\frac{1}{\sqrt{2\eta-1}}\left[x_\theta'-\sqrt{\frac{\eta}{2}}\left(\alpha e^{-i\theta}+\alpha^{\ast} e^{i\theta}\right)\right].
\label{ydef}
\end{equation}
Note that the derivation actually yields a complex $S_Q$. However, the imaginary part of $S_Q$ is so-called null function whose average over 
 all physically allowed quadrature distributions vanishes because the Q-function is real by definition. Therefore, the imaginary part of $S_Q$ can be safely neglected and it suffices to keep 
 only the real part as given in Eq. (\ref{SQformula}). 

\begin{figure}[!t!]
\centerline{\includegraphics[width=0.9\linewidth]{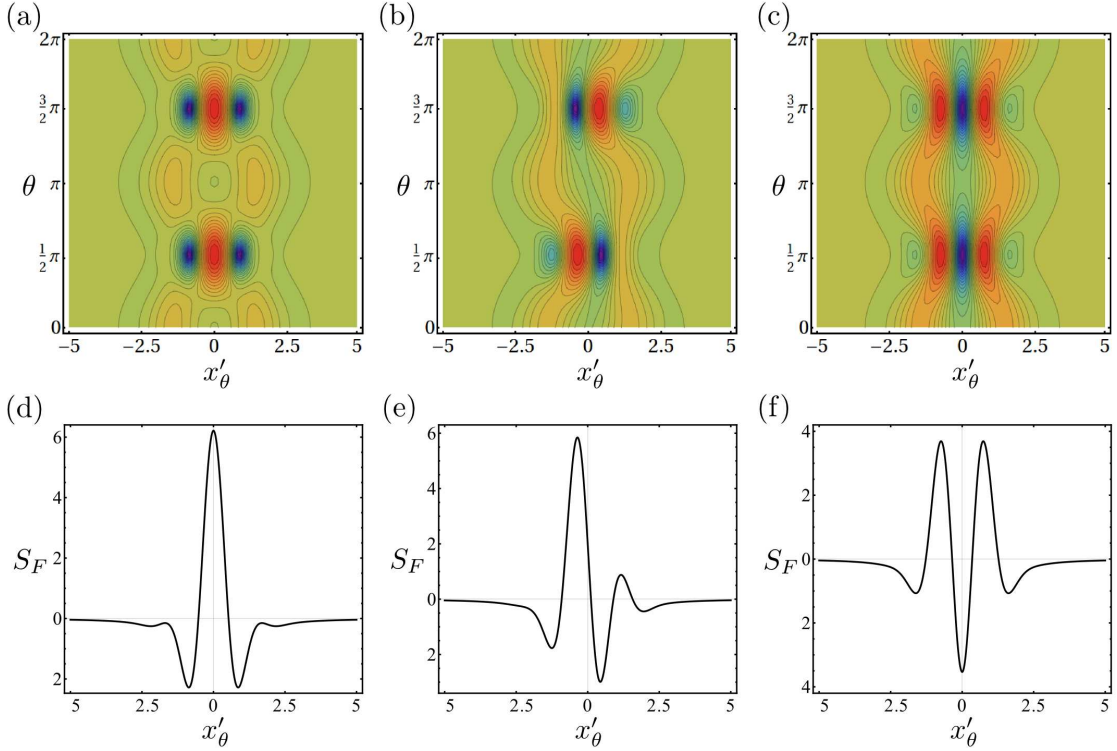}}
\caption{(Color online) Contour plots (a--c) and cross-sections at $\theta=\frac{\pi}{2}$ (d--f) of the sampling functions $S_F$ for the fidelity with cat-like state (\ref{catstate}) with fixed amplitude $\alpha=1$ and varying phase 
$\phi=0$ (panels a,d), $\phi=\frac{\pi}{2}$ (panels b,e), and $\phi=\pi$ (panels c,f). Perfect homodyne detection with $\eta=1$ is assumed.}
\end{figure}

 \begin{figure}[!t!]
\centerline{\includegraphics[width=0.9\linewidth]{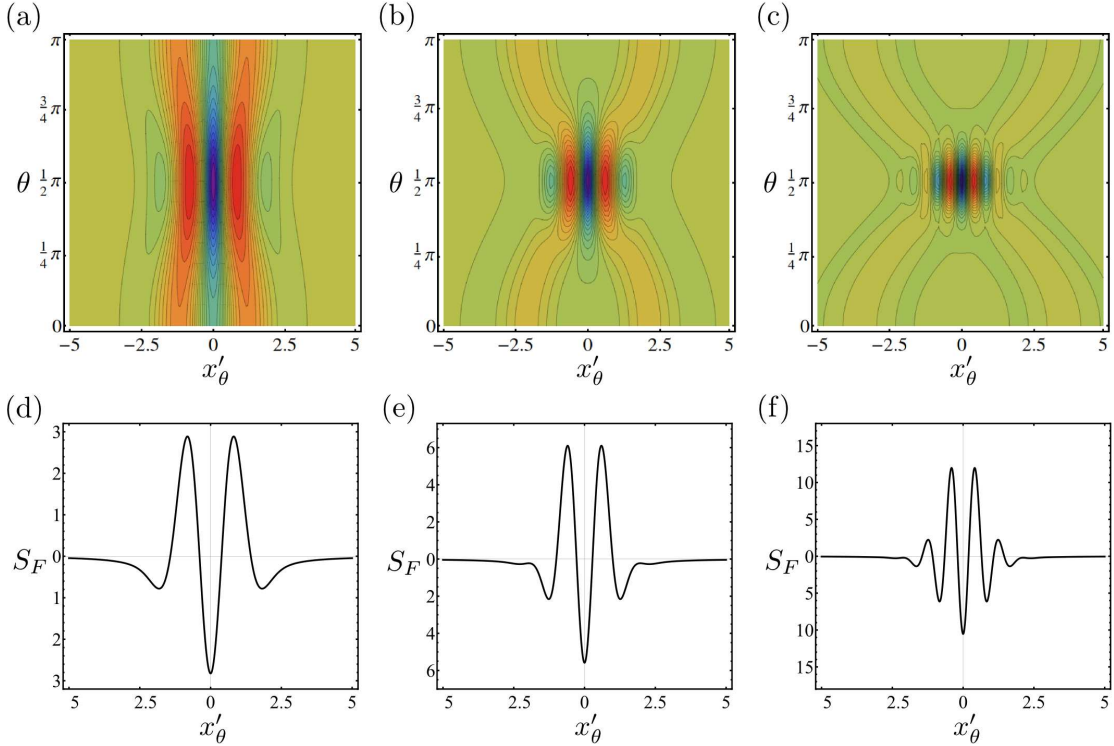}}
\caption{(Color online) Contour plots (a--c) and cross-sections at $\theta=\frac{\pi}{2}$ (d--f) of the sampling functions $S_F$ for the fidelity with odd coherent state ($\phi=\pi$) of three different amplitudes 
 $\alpha=0.75$ (panels a,d), $\alpha=1.5$ (panels b,e), and $\alpha=2.5$ (panels c,f). Perfect homodyne detection with $\eta=1$ is assumed.}
\end{figure}

With the sampling function for the Husimi Q-function at hand we can straightforwardly construct the sampling function for the fidelity with the cat-like state (\ref{catstate}).
It follows from Eq. (\ref{FCQ}) that
\begin{eqnarray}
S_F(x_\theta',\theta;\eta,\alpha,\phi)&=&\frac{\pi}{2\mathcal{N}}\left[S_Q(x_\theta',\theta;\eta,\alpha,\alpha^\ast)+S_Q(x_\theta',\theta;\eta,-\alpha,-\alpha^\ast)\right] \nonumber  \\
& & +\frac{\pi}{2\mathcal{N}}e^{-2|\alpha|^2}\left[e^{i\phi}S_Q(x_\theta',\theta;\eta,-\alpha,\alpha^\ast)+e^{-i\phi}S_Q(x_\theta',\theta;\eta,\alpha,-\alpha^\ast)\right].
\end{eqnarray}
Note that the explicit determination of $S_F$ requires evaluation of the error function of a complex argument. 
We recall that due to the relation $x_{\theta+\pi}=-x_\theta$ it actually suffices to sample the rotated quadratures
only in the interval $\theta\in[0,\pi]$. In fact, the sampling function exhibits the symmetry $S_F(x_\theta,\theta)=S(-x_\theta,\theta+\pi)$.
This symmetry is illustrated in Fig. 1 where the sampling function $S_F$ is plotted for three different values of the phase $\phi$ and a fixed amplitude $\alpha$.
The dependence of the sampling function $S_F$ on $x_\theta$ exhibits oscillatory  behaviour that reflects the phase-space interference 
characteristic for the cat-like states (\ref{catstate}). As shown in Fig.~2, the number of visible interference fringes grows with the increasing amplitude $\alpha$.
The loss-compensating sampling function can be obtained by a suitable re-scaling of the sampling function for perfect homodyne detection with $\eta=1$. According to Eq. (\ref{ydef}),
we have to re-scale the measured quadrature values as $x_\theta'\rightarrow x_{\theta}'/\sqrt{2\eta-1}$ and we have to increase the effective cat-state amplitude, 
$\alpha \rightarrow \sqrt{\frac{\eta}{2\eta-1}}\alpha$.
Moreover, the whole sampling function is multiplied by a factor $\frac{\eta}{2\eta-1}>1$.  
 The inefficient detection can be compensated only for $\eta>0.5$ and the sampling function $S_F$ becomes divergent when $\eta\rightarrow 0.5$.

The fidelity $F_{-}=\langle \alpha_{-}|\rho|\alpha_{-}\rangle$ with  an odd coherent state 
\begin{eqnarray}
|\alpha_{-}\rangle&=&\frac{1}{\sqrt{2-2e^{-2|\alpha|^2}}}(|\alpha\rangle-|-\alpha\rangle) 
\end{eqnarray}
is of particular interest because it represents a witness of the negativity of Wigner function at the origin of phase space. 
Recall that a quantum witness can be defined as an operator $A$ chosen such that
if $\mathrm{Tr}[\rho A]>T_A$, where $T_A$ is a certain threshold, then the state $\rho$ exhibits the witnessed property. The entanglement  witnesses \cite{Terhal00,Lewenstein00,Lewenstein01,Barbieri03} and non-classicality witnesses \cite{Shchukin05,Korbicz05,Kiesel12,Mari11} were extensively studied in the literature and recently also witnesses of quantum non-Gaussianity 
were introduced \cite{Filip11,Jezek11,Jezek12,Genoni13}. With the help of quantum witnesses one can experimentally verify and certify important features of a quantum state without full reconstruction of the studied quantum state.

The Wigner function at the origin is proportional to the mean value of photon number parity operator,
 $W(0,0)=\frac{1}{\pi}\langle (-1)^{n}\rangle$.  Let $\Pi_{-}$ and $\Pi_{+}$ denote the projectors onto subspaces spanned by odd and even Fock states, respectively.
It holds that $\Pi_{-}+\Pi_{+}=\mathbb{I}$, where $\mathbb{I}$ denotes the identity operator. The state $|\alpha_{-}\rangle$ belongs to the subspace spanned by the odd Fock states,
 $\Pi_{-}|\alpha_{-}\rangle=|\alpha_{-}\rangle$, which implies that $\Pi_{-}-|\alpha_{-}\rangle\langle \alpha_{-}| \geq 0$. Consequently, it holds for any state $\rho$ that
\begin{equation}
\mathrm{Tr}[\Pi_{-}\rho] \geq F_{-}.
\end{equation}
With the help of this inequality we can upper bound the value of Wigner function at the origin of phase space as follows,
\begin{eqnarray}
W(0,0) &=&-\frac{1}{\pi}\mathrm{Tr}[\Pi_{-}\rho]+\frac{1}{\pi} \mathrm{Tr}[\Pi_{+}\rho]  
= -\frac{1}{\pi}\mathrm{Tr}[\Pi_{-}\rho]+\frac{1}{\pi}[1-\mathrm{Tr}(\Pi_{-}\rho)]  \nonumber  \\
&=& \frac{1-2\mathrm{Tr}[\Pi_{-}\rho]}{\pi} \leq \frac{1-2F_{-}}{\pi}.
\label{W0bound}
\end{eqnarray}
In particular, if $F_{-}>\frac{1}{2}$, then the Wigner function is negative at the origin, $W(0,0)<0$.
The fidelity $F_{-}$ thus provides a lower bound on the negativity of Wigner function and its estimation allows us to certify
the negativity even without full tomographic reconstruction of the Wigner function. 

Very recently, it was shown that the value of Wigner function at the origin of phase space 
provides a witness of a quantum non-Gaussian character of the studied state \cite{Genoni13}. A quantum state is quantum non-Gaussian if and only if it cannot be expressed as a convex 
mixture of Gaussian states \cite{Filip11,Jezek11,Genoni13}.
It was proved in Ref. \cite{Genoni13} that the state is quantum non-Gaussian if 
\begin{equation}
W(0,0)< \frac{1}{\pi} \exp[-2\bar{n}(\bar{n}+1)].
\label{Wnongaussian}
\end{equation}
 Here $\bar{n}$ denotes the mean number of photons which can be estimated by averaging a sampling function $f_{\bar{n}}=\left[(x_\theta')^2-\frac{1}{2}\right]/\eta$ over the homodyne data \cite{Richter99jmo}. 
 By combining the inequalities (\ref{W0bound}) and (\ref{Wnongaussian}) we find that the fidelity with an odd coherent state can also serve as a witness of the quantum non-Gaussianity.
 In particular, the state is quantum non-Gaussian if
 \begin{equation}
 F_{-}> \frac{1}{2}- \frac{1}{2}\exp[-2\bar{n}(\bar{n}+1)],
 \label{Fqng}
 \end{equation}
 Note that the right-hand side of this inequality is smaller than $\frac{1}{2}$ for any finite $\bar{n}$.
 The class of non-Gaussian states is strictly larger than class of states with negative Wigner function, because there exist quantum non-Gaussian states with positive Wigner function.
By proving that the state is quantum non-Gaussian we certify that a highly non-linear process was involved in its preparation \cite{Jezek12,Genoni13}.

\section{Sampling functions for fidelities with squeezed Fock states}
In practice, an approximate version of the cat-like state (\ref{catstate}) can be prepared by subtracting a single photon from a squeezed vacuum state \cite{Ourjoumtsev06,Neergaard06,Takahashi08,Gerrits10}. Under ideal circumstances, the resulting state coincides with the pure squeezed single-photon state,
\begin{equation}
|\psi_{S}\rangle=U(r)|1\rangle,
\end{equation}
where $U(r)=e^{-ir(xp+px)/2}$ denotes the unitary squeezing operator and $r$ is the squeezing constant. The fidelity of squeezed single-photon with an odd coherent state \cite{Lund04}
\begin{equation}
F(\alpha,r)=\frac{2|\alpha|^2 \exp[-|\alpha|^2(1-\tanh r)]}{(\cosh r)^3 [1-\exp(-2|\alpha|^2)]}
\end{equation}
can be maximized by optimizing the squeezing $r$ as a function of $\alpha$. For $|\alpha| \leq 1$ the maximum achievable fidelity exceeds $0.997$, hence the state $U(r)|1\rangle$
represents an excellent approximation to the odd coherent state $|\alpha_{-}\rangle$. Since the Fock-state expansion of a squeezed single-photon state
contains only odd Fock states, the fidelity with this state can serve as a witness of Wigner function negativity similarly to the fidelity with an exact odd coherent state.

In this  section we derive sampling functions for state fidelity with squeezed Fock states. In other words, we seek sampling functions $f_{\mathrm{sq},n}(x_\theta,\theta;\eta,r)$ 
for diagonal density matrix elements in the basis of squeezed Fock states, $p_n(r)=\langle n|U^\dagger(r)\rho U(r)|n\rangle$,
\begin{equation}
p_n(r)=\frac{1}{\pi}\int_0^\pi \int_{-\infty}^{\infty} f_{\mathrm{sq},n}(x'_\theta,\theta;\eta,r) \, w(x'_\theta;\theta,\eta) \, d x'_\theta \,d\theta.
\label{pnrestimate}
\end{equation} 
We will make use of the fact that the Husimi Q-function is a generating function of density matrix elements in Fock basis,
\begin{equation}
\rho_{m,n}= \left.\frac{\pi}{\sqrt{m!\,n!}} \frac{\partial^{m+n}}{\partial\alpha^{\ast m }\partial \alpha^{n}} \left[e^{|\alpha|^2}Q(\alpha,\alpha^\ast)\right] \right|_{\alpha=\alpha^\ast=0}.
\end{equation}
Consequently, the sampling function for Husimi Q-function is a generating function for the sampling functions of density matrix elements in Fock basis. 
We will therefore first derive sampling function $S_Q(x_\theta,\theta;\eta,\alpha,\alpha^\ast,r)$ for Husimi Q-function of a squeezed version of the measured state $U^\dagger(r)\rho U(r)$. 
Subsequently, we will determine the sampling functions for $p_n(r)$ according to the formula
\begin{equation}
f_{\mathrm{sq},n}(x_\theta;\theta;\eta,r)=\left.\frac{\pi}{n!} \frac{\partial^{2n}}{\partial \alpha^n \partial \alpha^{\ast n}} \, e^{|\alpha|^2}S_Q(x_\theta,\theta;\eta,\alpha,\alpha^\ast,r) \right|_{\alpha=\alpha^\ast=0}.
\label{fgenerating}
\end{equation}

We proceed by expressing the measured rotated quadrature $x_\theta^\prime$ in terms of the quadrature operators $x_0=x e^{-r}$ and $p_0=p e^{r}$ of the squeezed state,
\begin{equation}
x_{\theta}^\prime= \sqrt{\eta}\left(x_0 e^r \cos\theta+p_0 e^{-r} \sin\theta\right)+\sqrt{1-\eta}x_{\mathrm{vac}}.
\end{equation}
We can rewrite this expression as follows,
\begin{equation}
x_{\theta}^\prime=\sqrt{\eta}\, a \,x_{0,\vartheta}+\sqrt{1-\eta} \,x_{\mathrm{vac}}. 
\label{xtheta}
\end{equation}
where $x_{0,\vartheta}=x_0\cos\vartheta+p_0\sin\vartheta$, and the new effective phase $\vartheta$ and the scaling factor $a$ are given by \cite{Jezek12}
\begin{equation}
\tan\vartheta=e^{-2r}\tan\theta, 
\label{vartheta}
\end{equation}
\begin{equation}
a=\sqrt{e^{2r}\cos^2\theta+e^{-2r}\sin^2\theta}.
\label{adef}
\end{equation}
With the help of Eq. (\ref{xtheta}) we can establish the following relation between characteristic functions of quadrature distributions,
\begin{equation}
 \left \langle \exp\left(\sqrt{2} i k x_{0,\vartheta}\right)\right\rangle =
  \exp\left(\frac{1-\eta}{2\eta}\frac{k^2}{a^2} \right) \left \langle \exp\left(\frac{\sqrt{2}i k}{a\sqrt{\eta}} x_\theta^\prime\right)\right\rangle,
 \label{characteristic}
\end{equation}
which generalizes the formula (\ref{charfunctionrelation}).
If we combine Eqs. (\ref{Qfunction}) and (\ref{characteristic}) then we arrive at the expression for Husimi Q-function of squeezed state $U^\dagger(r)\rho U(r)$,
\begin{eqnarray}
Q(\alpha,\alpha^\ast;r)= \frac{2}{\pi^2}\,\Re \int_0^{\pi} \int_0^\infty k e^{-k^2/2}
e^{\frac{1-\eta}{2\eta a^2}k^2 } 
 \left \langle \exp\left(\frac{\sqrt{2} \, i k}{a\sqrt{\eta}} x_\theta^\prime\right)\right\rangle 
 e^{-\sqrt{2} \,i k \tilde{x}_\vartheta} \,d k \,d \vartheta,
 \label{Qlong}
\end{eqnarray}
where $\tilde{x}_\vartheta=\frac{1}{\sqrt{2}}(\alpha e^{-i\vartheta}+\alpha^\ast e^{i\vartheta})$ and $\Re(z)$ denotes the real part of a complex variable $z$.
Note that we have exploited the symmetry properties under the $\pi$ phase shift of $\vartheta$ to reduce the integration 
 interval over $\vartheta$  from $[0,2\pi]$ to $[0,\pi]$. In particular, the shift $\vartheta\rightarrow \vartheta+\pi$ induces the following transformations,
 $\theta\rightarrow \theta+\pi$, $x_{\theta}'\rightarrow -x_\theta'$, and $\tilde{x}_\vartheta\rightarrow -\tilde{x}_\vartheta$. 
 The integration over $\vartheta$ can be replaced with the integration over the actual phase $\theta$ between the local oscillator and the signal beam,
\begin{equation}
\int_{0}^{\pi}  d \vartheta= \int_{0}^{\pi} \frac{\rm d \vartheta}{d \theta} \, d \theta =\int_{0}^{\pi} \frac{1}{a^2} \, d \theta,
\end{equation}
and after this substitution  we can repeat the procedure outlined in Section II. We express the quadrature characteristic function appearing in Eq. (\ref{Qlong})  
in terms of the measured quadrature distribution, exchange the order of integrals and integrate over $k$.
After some algebra we obtain sampling function for Husimi Q-function of squeezed version of the measured state,
\begin{equation}
S(x_\theta',\theta;\eta,\alpha,\alpha^\ast,r)=\frac{2}{\pi a^2 s}\left[1-\sqrt{\pi}ye^{-y^2}\mathrm{erfi}(y)\right],
\label{SQr}
\end{equation}
where
\begin{equation}
s=\frac{\eta(a^2+1)-1}{\eta a^2}, \qquad 
y=\frac{1}{\sqrt{s}}\left(\frac{x_\theta'}{a\sqrt{\eta}}-\tilde{x}_\vartheta\right).
\end{equation}
The sampling function exists if and only if $s>0$ for all $\theta$ which implies a lower bound on the detection efficiency,
\begin{equation}
\eta > \frac{1}{1+e^{-2r}}.
\end{equation}
The amount of tolerable losses thus decreases with increasing squeezing.
\begin{figure}[!t!]
\centerline{\includegraphics[width=0.9\linewidth]{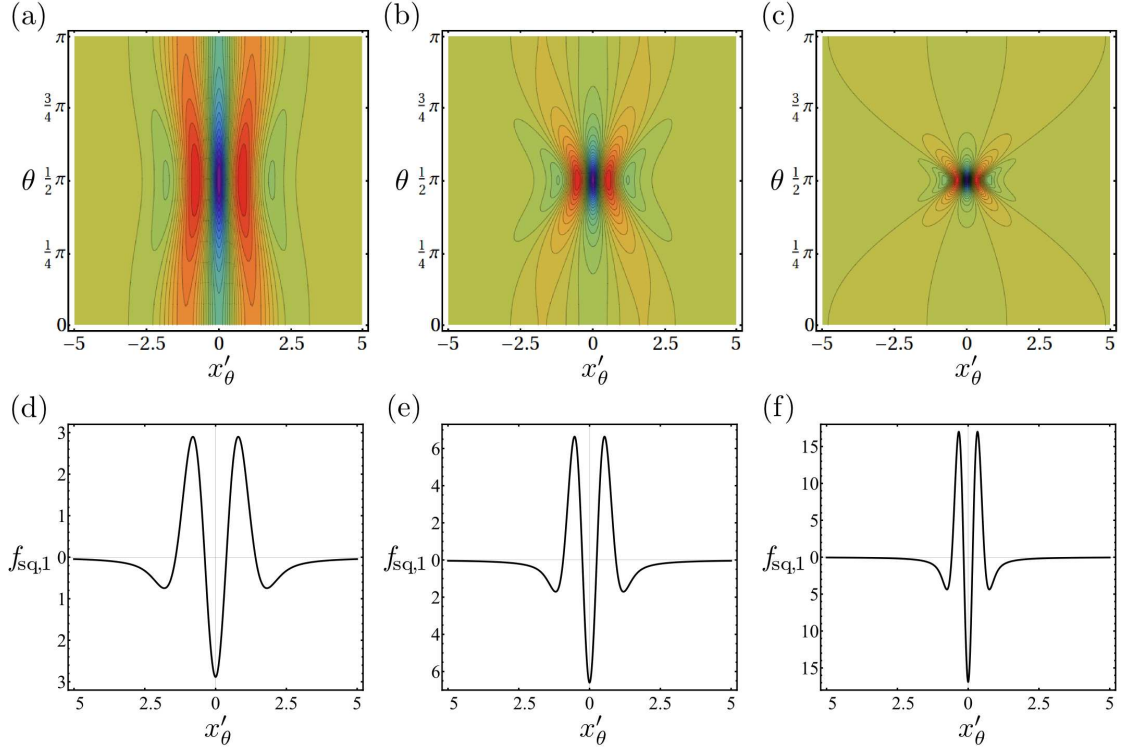}}
\caption{(Color online) Contour plots (a--c) and cross-sections at $\theta=\frac{\pi}{2}$ (d--f) of the sampling functions $f_{\mathrm{sq},1}$ for the fidelity with squeezed single-photon state with three different amounts of squeezing
 $r=0.183$ (panels a,d), $r=0.597$ (panels b,e), and $r=1.067$ (panels c,f). Perfect homodyne detection with $\eta=1$ is assumed.}
\end{figure}
We now insert the sampling function (\ref{SQr}) into Eq. (\ref{fgenerating}) and calculate sampling functions for estimation of the fidelities with squeezed vacuum and squeezed single-photon states.
After some algebra we obtain 
\begin{eqnarray}
f_{\mathrm{sq},0}&=&\frac{1}{a^2 s}f_0(x_{\theta,\eta,r}), \nonumber \\
f_{\mathrm{sq},1}&=&\frac{1}{a^2s^2}\left[f_1(x_{\theta,\eta,r})-(1-s)f_0(x_{\theta,\eta,r})\right],
\end{eqnarray}
where \begin{equation}
x_{\theta,\eta,r}=\frac{x_\theta'}{\sqrt{\eta(a^2+1)-1}}
\end{equation}
and $f_n(x)$ denote the standard sampling functions for diagonal density matrix elements in Fock basis \cite{DAriano95,Leonhardt95a,Leonhardt95b,Richter00},
\begin{eqnarray}
f_0(x)&=&2 - 2  \sqrt{\pi} x e^{-x^2} \mathrm{erfi}(x),\nonumber \\
f_1(x)&=&2(2x^2-1) + 4 \sqrt{\pi} x (1-x^2)e^{-x^2} \mathrm{erfi}(x).
\label{f01simplest}
\end{eqnarray}
From these general formulas we can recover the previously derived pattern functions as special cases. If we set $r=0$
then $a=1$ and $s=(2\eta-1)/2$ and we recover the loss-compensating sampling functions for probability of vacuum a single-photon states. 
If $\eta=1$, then $s=1$ for any $a$ and the sampling functions for density matrix elements in squeezed Fock basis are obtained by a simple re-scaling of
sampling functions $f_n(x)$ for density matrix elements in the ordinary Fock basis \cite{Jezek12},
\begin{equation}
f_{\mathrm{sq},n}(x_\theta;\theta,r)=\frac{1}{a^2}f_n\left(\frac{x_\theta}{a}\right).
\label{fsqn}
\end{equation} 
In Fig. 3 we plot the sampling function $f_{\mathrm{sq},1}$ for three different values of $r$. The values of squeezing constant were chosen such as to maximize the fidelity $F(r,\alpha)$ 
between squeezed single-photon state and an odd coherent state with amplitudes $\alpha_1=0.75$,  $\alpha_1=1.5$, and $\alpha_1=2.5$, respectively. By comparing Figs. 2 and 3 we can see that
for $\alpha=0.75$ the sampling functions $f_{\mathrm{sq},1}$ and $S_{F}$ practically coincide because the fidelity between the odd coherent state and squeezed single-photon state is extremely high.
The similarity  between the two sampling functions reduces with increasing cat-state amplitude $\alpha$ and for $\alpha=2.5$ the differences already become very significant.

\section{Reduction of statistical uncertainty by optimized quadrature sampling}
In the experiments, the phase $\theta$ is usually either set to several equidistant fixed values in the interval $[0,\pi]$, or it is linearly varied with time such that
the number of quadrature samples does not depend on $\theta$. However, this strategy generally may not be optimal in a sense that a smarter choice of the 
dependence of the number of samples $m(\theta)\,d \theta$ on $\theta$ may lead to reduction of the statistical uncertainty of the estimated fidelity. 
If the state is squeezed, then one may intuitively expect that it could be beneficial to collect more samples of the squeezed quadratures where one needs to resolve finer features
than for the anti-squeezed quadrature. Here we will confirm this intuition with exact analytical calculations. 
Let $M=\int_0^\pi m(\theta)\,  d \theta$ denote the total number of quadrature measurements. We will seek the optimal $m(\theta)\geq 0$ that for a given fixed $M$ minimizes the variance 
$V$ of an estimate obtained by averaging some sampling function $f(x_\theta;\theta)$ over the measured quadrature statistics $w(x_\theta;\theta)$. 
Since all results of quadrature measurements are mutually independent random variables, the variance can be expressed as \cite{DAriano99}
\begin{equation}
V=\frac{1}{\pi^2}\int_{0}^{\pi} \frac{1}{m(\theta)}V_f(\theta) \,d \theta,
\label{Vdef}
\end{equation}
where
\begin{equation}
V_{f}(\theta)= \int_{-\infty}^{\infty} f^2(x_\theta;\theta) w(x_\theta;\theta) \,d x_\theta - \left(  \int_{-\infty}^{\infty} f(x_\theta;\theta,r) w(x_\theta;\theta) \,d x_\theta\right)^2.
\label{Vf}
\end{equation}
The optimal sampling density that minimizes $V$ for a fixed $M$ can be found  with the help of the calculus of variations and reads
\begin{equation}
m_{\mathrm{opt}}(\theta)=\frac{M}{\int_0^\pi\sqrt{V_f(\phi)} \,d \phi }  \sqrt{V_f(\theta)}.
\label{moptdef}
\end{equation}
The corresponding minimum achievable variance is given by
\begin{equation}
V_{\mathrm{min}}=\frac{1}{M} \left[\frac{1}{\pi}  \int_0^\pi\sqrt{V_f(\theta)} \,d \theta\right]^2.
\end{equation}
It is instructive to compare this variance with the variance for a constant sampling density $m=M/\pi$, 
\begin{equation}
V_{\mathrm{c}}=\frac{1}{M\pi}\int_0^\pi V_f(\theta)\, d\theta .
\end{equation}
It holds that  $V_{\mathrm{min}}\leq V_{\mathrm{c}}$ due to Cauchy-Schwarz inequality and the equality holds only if $V_f(\theta)$ is constant.

As an illustrative example, let us consider estimation of diagonal density matrix elements in squeezed Fock basis. We shall assume that the measured state $\rho$ is a squeezed thermal-like state, $\rho=U(r)\rho_{T}U^\dagger(r)$, where the density matrix $\rho_T$ is diagonal in Fock basis, $\rho_T=\sum_{n=0}^\infty p_{T,n}|n\rangle \langle n|$. The quadrature distribution of this state can be expressed as
\begin{equation}
w(x_\theta;\theta)=\frac{1}{a} w_T\left(\frac{x_\theta}{a}\right),
\label{wT}
\end{equation}
where $w_T(x)$ is a distribution of an arbitrary quadrature of the state $\rho_{T}$ and for the sake of simplicity we assume perfect homodyning with unit detection efficiency. If we insert the quadrature distribution (\ref{wT}) and sampling functions (\ref{fsqn}) into Eq. (\ref{Vf}) we obtain
\begin{equation}
V_{f_{\mathrm{sq},n}}(\theta)=\frac{1}{a^4} V_{n},
\label{Vfthetan}
\end{equation}
where $a$ depends on $\theta$ according to Eq. (\ref{adef}) and
\begin{equation}
V_{n}=\int_{-\infty}^{\infty} f_{n}^2(x) w_T(x) \,d x - \left(  \int_{-\infty}^{\infty} f_{n}(x) w_T(x) \,d x\right)^2.
\end{equation}
is a constant that does not depend on $\theta$.
Let us first consider the homogeneous sampling with $m(\theta)=M/\pi$. The integral in Eq. (\ref{Vdef}) can be analytically evaluated and the variance of the 
estimate of $p_n(r)$ is given by
\begin{equation}
V_{\mathrm{sq},n}=\frac{1}{M}\,V_{n} \cosh(2r).
\end{equation}
The variance thus exponentially increases with the squeezing constant $r$. The optimal dependence of the sampling density $m(\theta)$ on $\theta$ is given by 
\begin{equation}
m_{\mathrm{opt}}(\theta)=\frac{1}{a^2}\frac{M}{\pi}
\end{equation}
and $m_\mathrm{opt}(\theta)$ does not depend on $n$ for the studied example.
It can be easily verified that $\int_0^\pi m_{\mathrm{opt}}(\theta)\, d \theta=M$. This optimal sampling density yields the variance 
\begin{equation}
V_{\mathrm{sq},n, \mathrm{min}}=\frac{1}{M}\,V_{n},
\end{equation}
which represents an improvement by a factor of $\cosh(2r)$ with respect to the homogeneous sampling. As anticipated, the sampling density is highest for 
the squeezed quadrature ($\theta=\pi/2$) and lowest for the anti-squeezed quadrature ($\theta=0$) and the ratio of these two sampling densities 
scales as $e^{4r}$. 

A successful application of the above discussed noise reduction procedure requires the knowledge which quadrature is squeezed and what is the amount of squeezing.
In many experiments on preparation  of nonclassical states of light, such information would be a-priori available to some extent, e.g.
from auxiliary measurements on the Gaussian squeezed vacuum state before photon subtraction. As this latter conditional operation typically exhibits 
 very small success rate, the reduction of statistical uncertainty by optimized quadrature sampling may be particularly relevant in this context. 
Even if the state would be initially completely unknown, one can perform 
probe measurements on several copies of the state to characterize its squeezing properties  and then use this information to 
optimize the dependence of the number of measurements on the phase shift between signal and local oscillator beams.

\section{Conclusions}

In summary, we have derived analytical formulas for the sampling functions that enable direct estimation of fidelity of a quantum state with coherent superposition of two coherent states or with a squeezed Fock state. The fidelity is determined by averaging the sampling functions over the quadrature statistics measured for phases $\theta$ spanning the interval $[0,\pi]$. The sampling functions can compensate for losses and inefficient homodyne detection
provided that the overall detection efficiency exceeds certain threshold. We stress that any application of the loss compensating sampling functions
requires careful and conservative estimation of the actual detection efficiency because any over-compensation may lead to overly optimistic or
even to unphysical estimates. We have pointed out that the fidelity with an odd coherent state as well as the fidelities with squeezed odd Fock states
provide a witness of the negativity of Wigner function and a witness of quantum non-Gaussian character of the measured state.
 Finally, we have noted that the statistical uncertainty of the fidelity estimates obtained by the
sampling-functions approach may be reduced by a suitable choice of the dependence of the number of samples on the phase shift $\theta$. 
The fidelity represents a convenient and widely used characteristics of the quality of quantum states.  We thus anticipate potential 
applications of our results to various experiments requiring characterization of highly nonclassical states of light.

\acknowledgments
We gratefully acknowledge stimulating discussions with R. Filip and L. Mi\v{s}ta, Jr. This work was supported by the Czech Science Foundation
under grant No. P205/12/0577.

\end{document}